\documentclass[reprint,amsmath,amssymb,aps,prl]{revtex4-2}
\usepackage[dvipdfmx]{graphicx}
\usepackage{dcolumn}
\usepackage{bm}
\usepackage{amsmath,amssymb}

\usepackage{color}
\begin{document}

\title{Bond-weighted Tensor Renormalization Group}
\author{Daiki Adachi$^1$}
\author{Tsuyoshi Okubo$^{1,2}$}
\author{Synge Todo$^{1,3}$}
\affiliation{
  $^1$Department of Physics, University of Tokyo, Tokyo, 113-0033, Japan\\
  $^2$JST, PRESTO, Tokyo, 113-0033, Japan\\
  $^3$Institute for Solid State Physics, University of Tokyo, Kashiwa, 277-8581, Japan
}

\begin{abstract}
We propose an improved tensor renormalization group (TRG) algorithm, the bond-weighted TRG (BTRG). In BTRG, we generalize the conventional TRG by introducing bond weights on the edges of the tensor network. We show that BTRG outperforms the conventional TRG and the higher-order tensor renormalization group with the same bond dimension, while its computation time is almost the same as that of TRG. Furthermore, BTRG can have non-trivial fixed-point tensors at an optimal hyperparameter. We demonstrate that the singular value spectrum obtained by BTRG is invariant under the renormalization procedure in the case of the two-dimensional Ising model at the critical point. This property indicates that BTRG performs the tensor contraction with high accuracy while keeping the scale-invariant structure of tensors.
\end{abstract}

\maketitle

Since Onsager proved the existence of a phase transition in the two-dimensional Ising model in 1944~\cite{Onsager1944}, critical phenomena have been one of the central subjects in statistical physics and condensed matter physics.
However, since only a few models have exact solutions, we generally need to conduct numerical simulations or approximated analytical calculations to investigate phase transitions and critical properties observed in exotic statistical models and materials.
Among various numerical techniques, the numerical renormalization group method based on the tensor network representation has become popular recently after the tensor renormalization group (TRG) proposed by Levin and Nave in 2007~\cite{LevinN2007}.
In TRG, we represent the partition function as a tensor network and perform tensor contractions using the low-rank approximation based on the singular value decomposition (SVD).
This technique enables us to evaluate the partition function and related physical quantities quite accurately for exponentially large systems, which can be regarded virtually as in the thermodynamic limit.
TRG and its variants have been successfully used to investigate a wide range of classical and quantum many-body systems~\cite{LiGZRGS2010, ChenQCWZNX2011, DittrichE2012, EvenblyV2016, WangXCBX2014, YuXMLDZQCX2014, UedaON2014, GenzorGN2016, YangLZXM2016, KawauchiT2016}.

It has been widely realized, however, that TRG becomes less accurate near the critical points.
To improve the accuracy of the numerical renormalization group for critical systems, the researchers have developed several algorithms so far.
For example, the second renormalization group (SRG)~\cite{XieJCWX2009} considers the low-rank approximation of the local tensors by taking the effect of environment tensors into account.
It has been demonstrated that SRG successfully improves accuracy, though it is much more expensive computationally~\cite{ZhaoXCWCX2010}. 

Another example is the higher-order tensor renormalization group (HOTRG)~\cite{XieCQZYX2012}, which uses the higher-order singular value decomposition (HOSVD) instead of SVD.
The accuracy of HOTRG is higher than the conventional TRG but lower than SRG.
One can understand the reason for the increased accuracy of HOTRG than TRG because HOTRG applies the HOSVD low-rank approximation to a pair of tensors.
It might also be natural that the accuracy of SRG is higher than HOTRG because SRG performs the low-rank approximation under a larger approximate environment.

More recently, the accuracy of the tensor network algorithms is discussed from the viewpoint of short-range loop entanglement.
In TRG, the loop entanglement, which represents short-range correlations in the physical systems, remains and accumulates during the renormalization steps.
By eliminating the effects of the loop entanglement directly, several methods achieved higher accuracy even near the critical point~\cite{EvenblyV2015, YangGW2017, GuW2009, Evenbly2017, Harada2018, BalMHV2017, HauruDM2018, WangQZ2014, ZhaoXXI2016, Evenbly2018, Ferris2013}.

Although the proposed methods produce more accurate results than the original TRG, they require significantly higher computation cost at the expense.
If one can improve the accuracy without increasing the computation cost, it will expand the applicability of the tensor network algorithms to problems that were previously difficult to apply.
An example of such methods might be the anisotropic tensor renormalization group (ATRG), which is recently proposed as the real space renormalization group for hypercubic lattices in general dimensions~\cite{AdachiOT2020}.
In Ref.~\cite{AdachiOT2020}, the present authors showed that without increasing the leading computation costs, ATRG improves the accuracy for the case of the two-dimensional Ising model.

In this Letter, we propose another improved TRG algorithm, the bond-weighted tensor renormalization group (BTRG), which achieves much higher accuracy than the conventional TRG without increasing the computation cost.
In BTRG, we introduce bond weights on the edges of the tensor network.
We demonstrate that in the case of the two-dimensional Ising model, BTRG outperforms the conventional TRG and HOTRG.
Interestingly, the singular value spectrum at the critical point, obtained by BTRG with an optimal hyperparameter, is stable under the renormalization procedure.
This observation indicates that BTRG captures the correct scale-invariant property of renormalized tensors at the critical point~\cite{EvenblyV2015}. 

\begin{figure}[tbp]
  \begin{center}
    \includegraphics[clip, width=7cm]{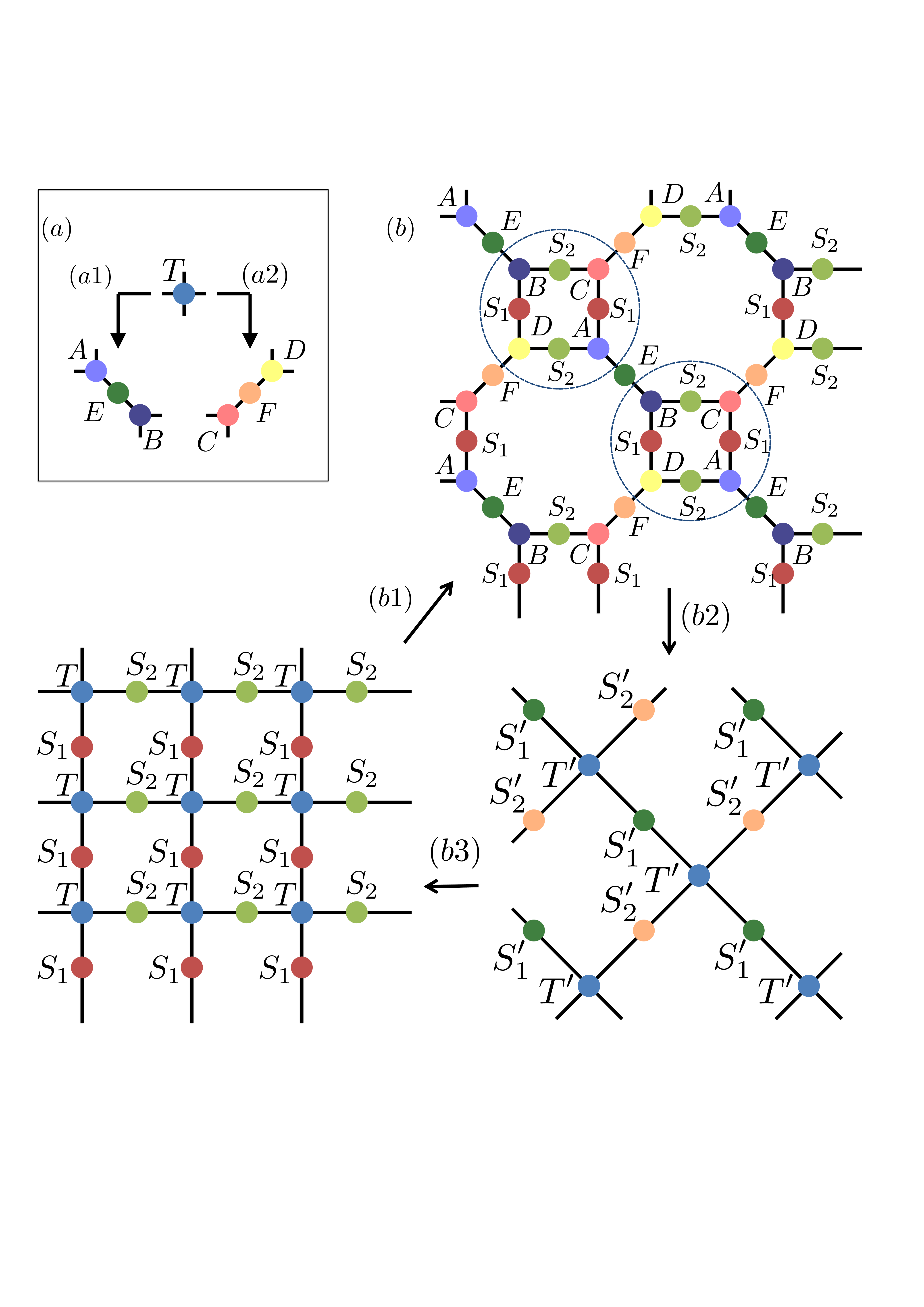}
  \end{center}
  \caption{\label{BTRG} (a)~Tensor decompositions in BTRG.
    A rank-4 site tensor ($T$) is decomposed into two rank-3 tensors ($A$ and $B$, or $C$ and $D$) and one rank-2 tensor ($E$ or $F$) depending on the position in the tensor network.
    Decompositions~(a1) and (a2) correspond to Eqs.~(\ref{TRG decomposition T1}) and (\ref{TRG decomposition T2}), respectively.
    (b)~Renormalization step of BTRG.
    (b1)~Tensor decomposition: Each rank-4 tensor is decomposed into two rank-3 tensors and one rank-2 tensor, according to (a1) or (a2).
    (b2)~Tensor contraction: Four rank-3 tensors ($A$, $B$, $C$, $D$) and four rank-2 tensors (two $S_1$'s and two $S_2$'s) are contracted into a new site tensor ($T'$), and the remaining tensors ($E$ and $F$) are regarded as new bond tensors ($S_1'$ and $S_2'$)~[Eqs.~(\ref{TRG contraction T})--(\ref{TRG contraction S2})].
    (b3)~Rescale: by rotating the new network by $\pi/4$ and rescaling by a factor $1/\sqrt{2}$, the original square-lattice structure is retained.
    In these network diagrams, the bond dimension of the legs (solid lines) is all $\chi$.
  }
\end{figure}

In BTRG, we consider renormalization of a tensor network with tensors locating not only on the vertices (sites) but also on the edges (bonds) (See Fig~\ref{BTRG}).
BTRG's renormalization process is slightly different from conventional TRG.
 In the original TRG, a rank-4 site tensor is decomposed into two rank-3 tensors. On the other hand, in BTRG, a rank-4 site tensor is decomposed into two rank-3 tensors and one rank-2 tensor, as shown in Fig.~\ref{BTRG}(a).

%{{{ Decomposition pictures in this paper
%}}}

Similar to the conventional TRG, first we apply the low-rank approximation to the 4-rank site tensor by using SVD. We introduce the following two different decompositions depending on the position in the tensor network [Fig.~\ref{BTRG}(a)]:
%{{{ TRG decomposition T1 & T2
\begin{align}
  \begin{split}
    T_{x_0, x_1, y_0, y_1} &\approx \sum_{i}^{\chi} {U_1}_{(x_0, y_0), i} {\sigma_1}_{ii} {V_1}_{i, (x_1, y_1)}, \label{TRG decomposition T1}
  \end{split} \\
  \begin{split}
    T_{x_0, x_1, y_0, y_1} &\approx \sum_{i}^{\chi} {U_2}_{(x_0, y_1), i} {\sigma_2}_{ii} {V_2}_{i, (x_1, y_0)}, \label{TRG decomposition T2}
  \end{split}
\end{align}
%}}}
where $\chi$ is the cutoff of the bond dimension.
Then, we define the tensors $A,B,C,D,E,F$ as
%{{{ TRG decompositions
\begin{align}
    A_{(x_0, y_0), i} &= {U_1}_{(x_0, y_0), i} {\sigma_1}_{ii}^{\frac{1-k}{2}}, \label{TRG tensor A} \\
    E_{i, j} &= \delta_{ij} {\sigma_1}_{ii}^{k}, \label{TRG tensor E} \\
    B_{i, (x_1, y_1)} &= {\sigma_1}_{ii}^{\frac{1-k}{2}} {V_1}_{i, (x_1, y_1)}, \label{TRG tensor B} \\
    C_{(x_0, y_1), i} &= {U_2}_{(x_0, y_1), i} {\sigma_2}_{ii}^{\frac{1-k}{2}}, \label{TRG tensor C} \\
    F_{i, j} &= \delta_{ij} {\sigma_2}_{ii}^{k}, \label{TRG tensor F} \\
    D_{i, (x_1, y_0)} &= {\sigma_2}_{ii}^{\frac{1-k}{2}} {V_2}_{i, (x_1, y_0)}. \label{TRG tensor D}
\end{align}
%}}}
Here, $k$ is a hyperparameter representing the difference from the original TRG. The present algorithm is reduced to the original TRG at $k=0$.
In the case of TRG ($k=0$), $E$ and $F$ [Eqs.~(\ref{TRG tensor E}) and (\ref{TRG tensor F})] are identity matrices, while for nonzero $k$, they contain information about the singular values.
After the decompositions of the rank-4 tensors, we create a new renormalized site tensors by contracting four rank-3 tensors and four rank-2 tensors, and regard $E$ and $F$ as new bond tensors as
%{{{ TRG contraction 2
\begin{align}
  \begin{split}
    T^{\prime}_{x_0, x_1, y_0, y_1} &= \sum_{i_0, i_1, i_2, i_3} \big[ B_{x_0, (i_0, i_2)} C_{(i_0, i_3), y_0} D_{y_1, (i_1, i_2)} \\
&\qquad \times A_{(i_1, i_3), x_1} {S_2}_{i_0, i_0} {S_2}_{i_1, i_1} {S_1}_{i_2, i_2} {S_1}_{_3, i_3} \big] , \label{TRG contraction T}
  \end{split} \\
  S_1^{\prime} &= E, \label{TRG contraction S1} \\
  S_2^{\prime} &= F. \label{TRG contraction S2}
\end{align}
%}}}
By rotating the new network by $\pi/4$ and rescaling by a factor $1/\sqrt{2}$, the original square-lattice structure is retained.
We present the whole renormalization step of BTRG in Fig.~\ref{BTRG}(b).
It is straightforward to confirm that the order of the costs of this algorithm is the same as the original TRG: it requires $O(\chi^5)$ computation cost and $O(\chi^3)$ memory footprints.

As an initial condition, we set the bond tensors $S_1$ and $S_2$ as identity matrices at the beginning of BTRG.
During the renormalization steps, they become non-trivial through the singular values $\sigma^{k}$ of the site tensors. The extra weights of the singular values $\sigma^{-k/2}$ are included in $A, B, C$, and $D$ tensors in addition to the weight $\sigma^{1/2}$ in the original TRG.
We may consider these weights as a mean-field environment similar to the mean-field SRG proposed in Ref.~\cite{ZhaoXCWCX2010}.
In the mean-filed SRG, one estimates the mean field by iterative calculations, while in BTRG, we use the singular values obtained in the previous step. Thus, no additional effort is required for estimating the environment. Nevertheless, BTRG with a proper choice of $k$ dramatically improves the accuracy, as shown below. 

%{{{ BTRG_result1
\begin{figure}
  \begin{center}
   \includegraphics[width=8cm]{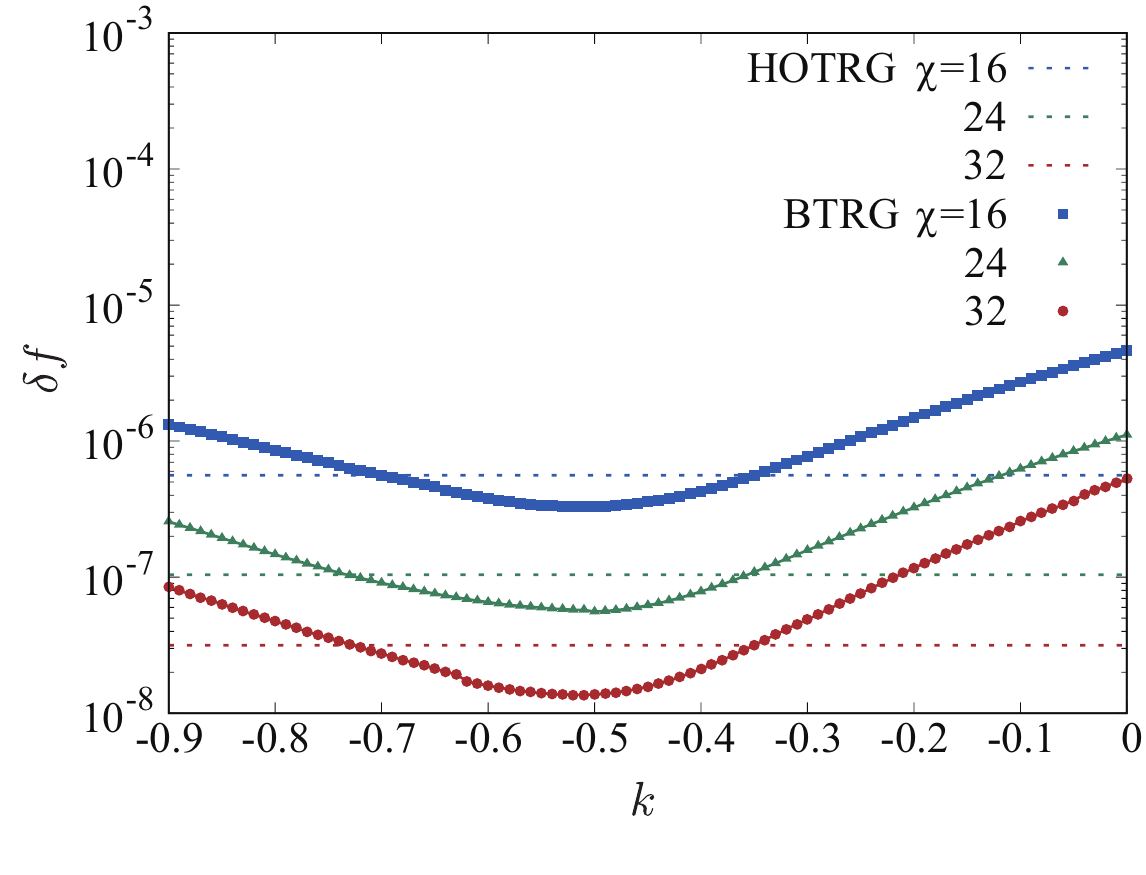}
  \end{center}
  \caption{$k$-dependence of the relative error of the free energy $ \delta f = \| f_{\text{calc}} - f_{\text{exact}}\|/\| f_{\text{exact}} \|$ by BTRG at the critical point $\beta=\beta_c$ with $\chi=16$ (blue squares), 24 (green triangles), and 32 (red circles).
    The results of HOTRG are also shown by the horizontal lines for comparison.}
  \label{BTRG_result1}
\end{figure}
%}}}

To demonstrate the advantage of BTRG, we calculate the two-dimensional Ising model on the square lattice by three different methods, TRG, HOTRG, and BTRG, and compare their results.
The initial site tensor is prepared in the same way as described in Ref.~\cite{XieCQZYX2012}.
The renormalization is performed 15 times for each axis for HOTRG and 30 times for TRG and BTRG.
First, we examine the $k$-dependence of the relative error of the free energy calculated by BTRG at the critical point $\beta_c = \frac{1}{2} \log \left(1 + \sqrt{2}\right)$.
As shown in Fig.~\ref{BTRG_result1}, by setting $k$ negative, we can reduce the relative error from TRG.
We do not show the results for $k>0$ in Fig.~\ref{BTRG_result1}, as the relative error becomes larger monotonically as $k$ increases.
One can see that BTRG gives the best result at $k \approx -0.5$.
Moreover, around the optimal point, the result of BTRG becomes more accurate than HOTRG with the same bond dimension $\chi$.
In the following calculation, we set $k$ as the optimal value, $k=-\frac{1}{2}$.

%{{{ BTRG_result3
\begin{figure}
  \begin{center}
    \includegraphics[width=8cm]{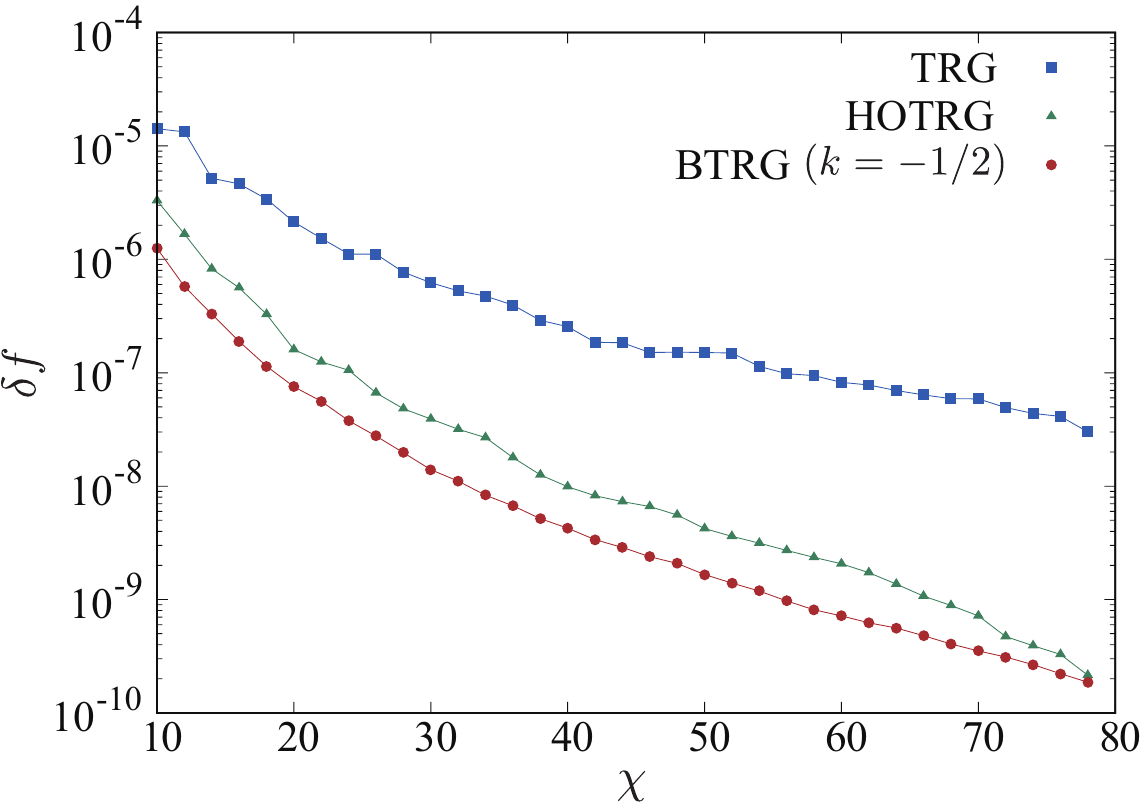}
  \end{center}
  \caption{
    $\chi$-dependence of the relative error of the free energy $ \delta f = \| f_{\text{calc}} - f_{\text{exact}}\|/\| f_{\text{exact}} \|$ by TRG (blue squares), HOTRG (green triangles), and BTRG (red circles) at the critical point $\beta=\beta_c$.
  }
  \label{BTRG_result3}
\end{figure}
%}}}

Next, we consider the $\chi$-dependence of the relative error of the free energy at the critical point (Fig.~\ref{BTRG_result3}).
The relative error of the free energy calculated by BTRG is smaller than those by TRG and HOTRG for all bond dimensions. 
%{{{ BTRG_result2
\begin{figure}
  \begin{center}
    \includegraphics[width=8cm]{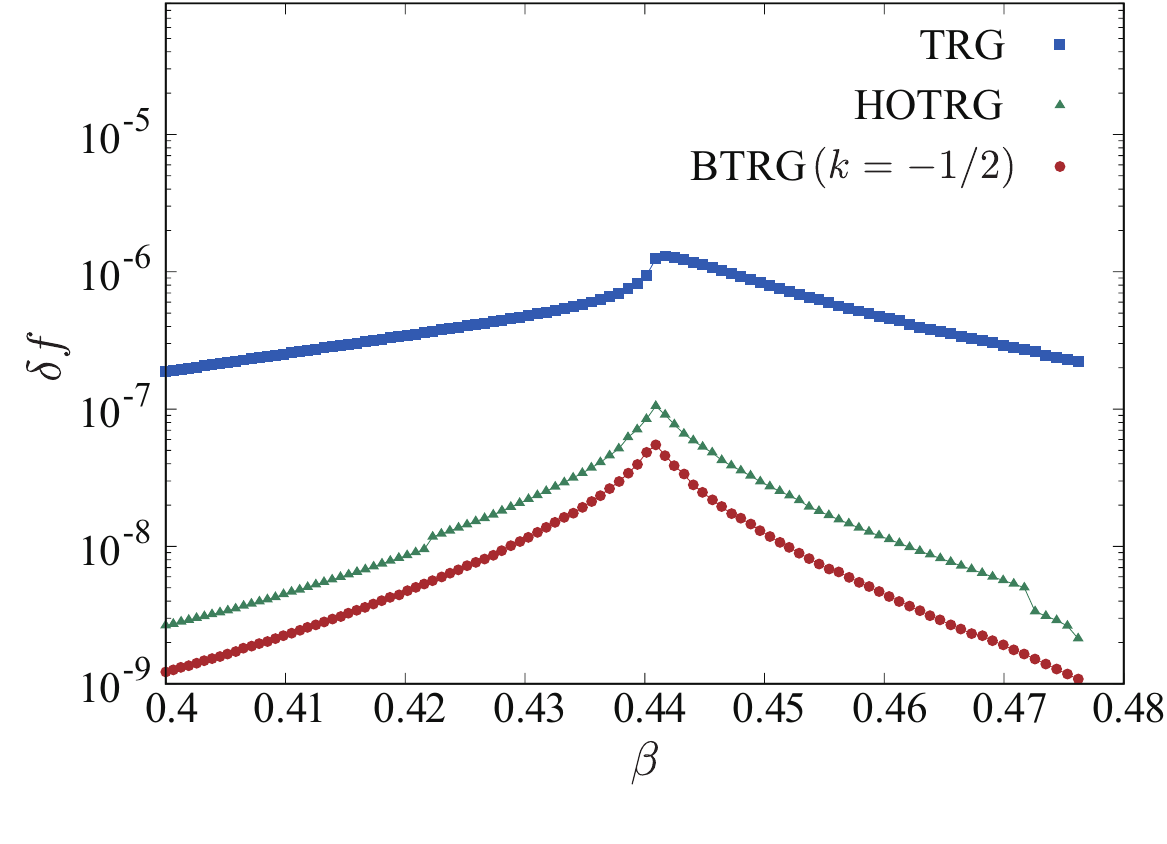}
  \end{center}
  \caption{
    $\beta$-dependence of the relative error of the free energy $ \delta f = \| f_{\text{calc}} - f_{\text{exact}}\|/\| f_{\text{exact}} \|$ by TRG (blue squares), HOTRG (green triangles), and BTRG (red circles) for $\chi=24$.
  }
  \label{BTRG_result2}
\end{figure}
%}}}
We also show the $\beta$-dependence of the relative error of the free energy in Fig.~\ref{BTRG_result2}. The bond dimension is $\chi=24$. 
Again, the relative error of BTRG is smaller than those by TRG and HOTRG at all temperatures.

We thus conclude that BTRG outperforms TRG and HOTRG for the square-lattice tensor networks from the above benchmark results.
It should be emphasized that the computation cost of BTRG is $O(\chi^5)$, which is the same as TRG and much smaller than that of HOTRG, $O(\chi^7)$.
Even though the improvement of BTRG over HOTRG is not that impressing with the same bond dimension $\chi$, the difference in the accuracy per computation cost becomes significant for large $\chi$.

%{{{ why k=-1/2 is the best
\begin{figure}
  \begin{center}
    \includegraphics[clip, width=5cm]{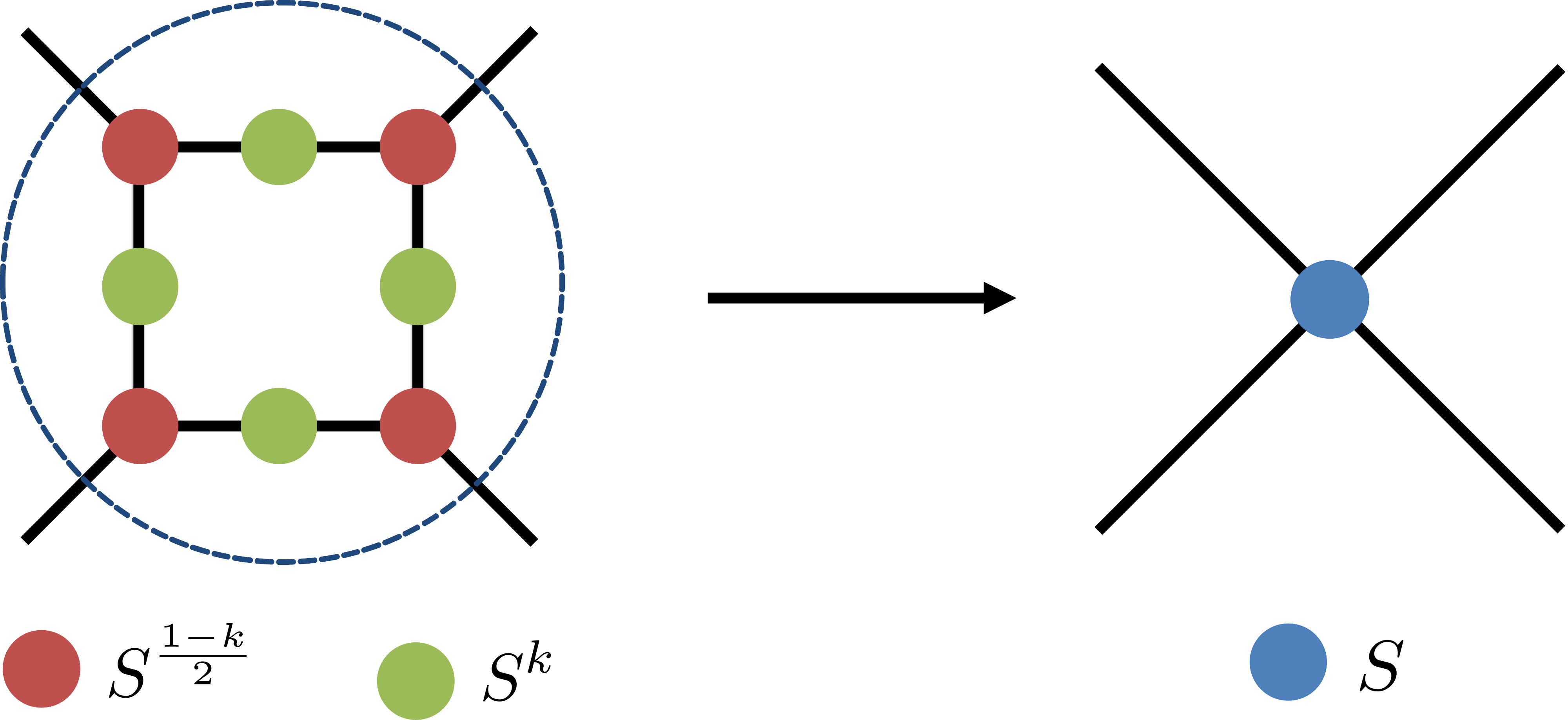}
  \end{center}
  \caption{
    Singular values of the fixed-point tensors in BTRG.
    After the tensor decomposition [Eqs.~(\ref{TRG decomposition T1})--(\ref{TRG tensor D})], we obtain rank-3 tensors with singular values $\sigma^{\frac{1-k}{2}}$ and rank-2 tensors with singular values $\sigma^{k}$.
    By contracting these tensors, a renormalized rank-4 tensor is created, which has the same singular values ($\sigma$) as the rank-4 tensor before the renormalization.
  }
  \label{TRG_bond3}
\end{figure}
%}}}

As discussed already, one may attribute the improvement of accuracy in BTRG to the effectively larger environment through the bond weights.
However, it is not trivial why $k \approx -\frac{1}{2}$ gives the optimal result.
Here, we discuss the reason by considering the stationary condition of the renormalization transformation.

We assume that the tensors $T$, $S_1$, and $S_2$ have already converged to some fixed-point tensors after several renormalization steps.
As a result, the singular value matrix $\sigma$ in Eqs.~(\ref{TRG decomposition T1}) and (\ref{TRG decomposition T2}) as well as the new bond tensors $E=F=\sigma^k$ in Eqs.~(\ref{TRG tensor E}) and (\ref{TRG tensor F}) become invariant.
Then, the renormalized site tensor $T'$ is created from four rank-3 tensors, which are proportional to $\sigma^{\frac{1-k}{2}}$, and four rank-2 tensors.
If we further assume that the unitary matrices do not contribute to the singular value spectrum, we have the following stationary condition:
%{{{ Singular values equation
\begin{align}
  \big[\sigma^{\frac{1-k}{2}}\big]^4 \big[\sigma^{k}\big]^4 &= \sigma.
  \label{Singular values equation}
\end{align}
%}}}
By solving this equation, we obtain $k=-\frac{1}{2}$ that gives a stable fixed point. 

Notice that when we assume the corner double line (CDL) structure for the fixed-point tensors, where each index consists of two lines representing the corner correlations, one of the two lines is absorbed as a normalization constant after one renormalization step. This fact indicates that only $\big[\sigma^{\frac{1-k}{2}}\big]^\frac{1}{2}$ from each tensor contributes to the singular value spectrum of the renormalized tensor. In this case, $k=0$ satisfies the stationary condition, which is consistent with the previous discussions that the CDL tensor is a fixed point in the conventional TRG~\cite{LevinN2007,GuW2009}. 

We can regard $k=-\frac{1}{2}$ as a necessary condition for the existence of a non-trivial stationary solution of BTRG. Interestingly, the condition $k=-\frac{1}{2}$ gives a scaling of the partition function $Z \propto \sigma^{0}$, which indicates the scale invariance of the free energy density. It is another evidence that supports $k= -\frac{1}{2}$ is optimal. 

%{{{ singularvaluse_for_comparison
\begin{figure}
 \begin{center}
  \includegraphics[width=8cm]{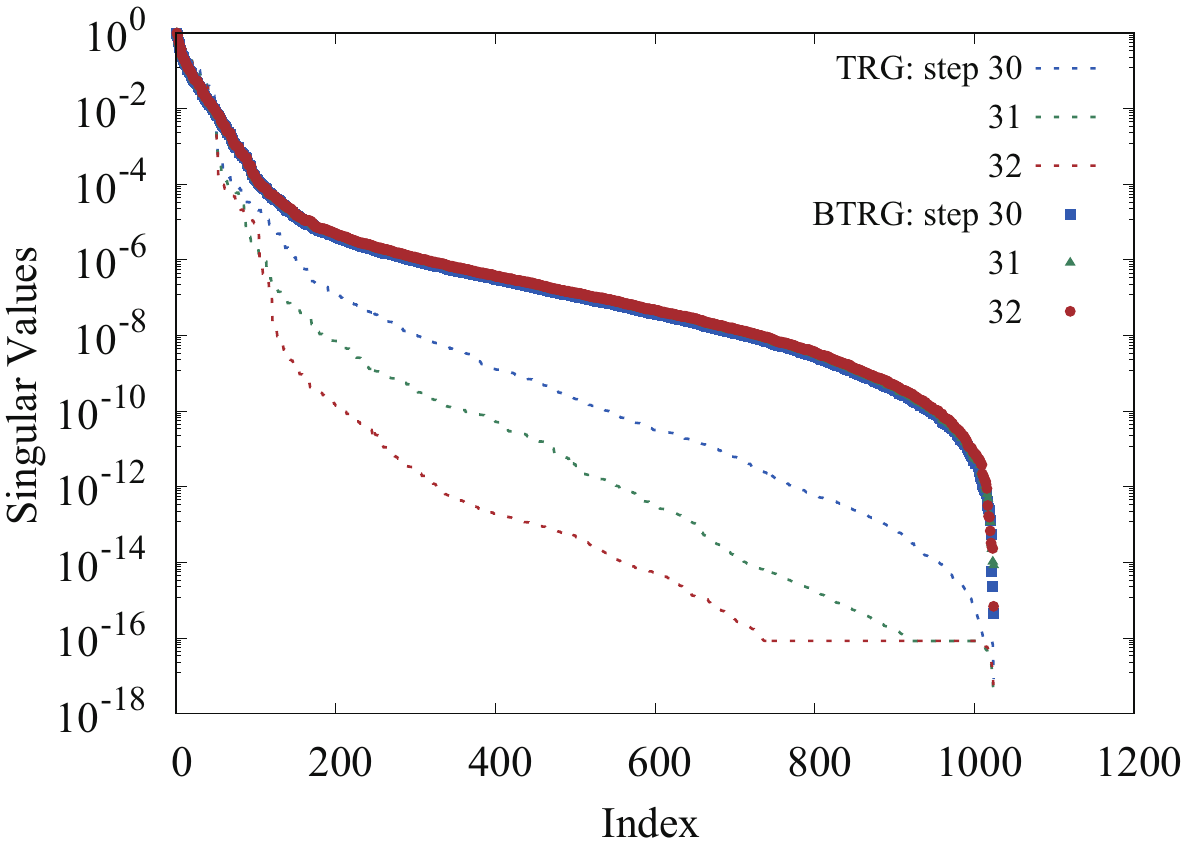}
 \end{center}
  \caption{  
    Singular value spectrum obtained by TRG (dashed lines) and BTRG (symbols) after 30, 31, and 32 renormalization steps.
    The bond dimension is $\chi=32$ in all the cases.
    The singular value spectrum obtained by TRG changes as the renormalization proceeds, while the spectrum obtained by BTRG does not exhibit any visible change in this scale.
  }
  \label{singularvalues_for_comparison}
\end{figure}
%}}}

To confirm the stability of the singular value spectrum, here, we show that singular values obtained in BTRG with $k=-\frac{1}{2}$ at the critical point of the two-dimensional Ising model.
First, we prepare the site tensor after 30, 31, and 32 iterations by TRG and BTRG with $\chi=32$.
Then, the singular value spectrum is calculated by Eq.~(\ref{TRG decomposition T1}). In Fig.~\ref{singularvalues_for_comparison},
we observe that the spectrum calculated by TRG changes as the renormalization proceeds, while that obtained by BTRG shows no significant change.
This result strongly supports the above argument that BTRG keeps the scale-invariant structure of tensors at the critical point.

The stability of the singular value spectrum also means the scaling dimensions calculated from the local tensors such as Ref.~\cite{GuW2009} are stable under the renormalization. Several works have indicated that removing the short-range entanglement is crucial to obtain the scaling dimensions stably~\cite{EvenblyV2015, YangGW2017, GuW2009, Evenbly2017, Harada2018, BalMHV2017, HauruDM2018, WangQZ2014, ZhaoXXI2016, Evenbly2018, Ferris2013}. However, by using BTRG, we can calculate scaling dimensions stably, without explicitly removing the short-range entanglement. It is another significant property of BTRG.

In this paper, we proposed a new tensor renormalization group method, BTRG. By considering a network where tensors locate on the sites and bonds, we generalized the original TRG. This approach requires $O(\chi^5)$ computation cost and $O(\chi^3)$ memory footprint, both of which are the same as TRG. Nevertheless, BTRG gives more accurate results with the optimal hyperparameter, $k=-\frac{1}{2}$, than not only TRG but also HOTRG. We showed that the stationary condition could explain the optimal hyperparameter $k=-\frac{1}{2}$ for non-trivial fixed-point tensors. Indeed, we confirmed that the singular value spectrum at the critical point obtained by BTRG with $k=-\frac{1}{2}$ becomes invariant under the renormalization.

In BTRG, by generalizing the tensor network to include bond tensors, site tensors naturally take the environment's effect into account at the decomposition procedure. It contributes to increasing the accuracy of the approximation. This idea can be easily applied to the other tensor renormalization groups, such as TRG on the honeycomb lattice and HOTRG, as shown in the supplementary materials \footnote{D. Adachi, T. Okubo, and S. Todo, Supplemental material for ``Bond-weighted Tensor Renormalization Group'' (SM.pdf)}. 

The nature of the fixed-point tensor in BTRG (with $k=-\frac{1}{2}$) is one of the essential issues to be investigated. As we have shown, the singular value spectrum becomes scale-invariant, indicating that it contains detailed critical properties. By understanding the structure of the fixed-point tensors, we might extract the scaling dimensions and fusion rules in the conformal field theory.

We thank K.~Harada, N.~Kawahima, and H.-H.~Zhao for fruitful discussion. D.~A. is supported by the Japan Society for the Promotion of Science through the Program for Leading Graduate Schools (MERIT).
This work is partially supported by MEXT as ``Exploratory Challenge on Post-K computer'' (Frontiers of Basic Science: Challenging the Limits) and ``Program for Promoting Researches on the Supercomputer Fugaku'' (Basic Science for Emergence and Functionality in Quantum Matter --Innovative Strongly-Correlated Electron Science by Integration of “Fugaku” and Frontier Experiments--), by JSPS KAKENHI No.~15K17701, 17K05564, 19K03740, and 20H00122, and by JST, PRESTO Grant Number JPMJPR1912, Japan.

\bibliography{BTRG}

\end{document}

\title{Supplemental material for ``Bond-weighted Tensor Renormalization Group''}
\author{Daiki Adachi$^1$}
\author{Tsuyoshi Okubo$^{1,2}$}
\author{Synge Todo$^{1,3}$}
\affiliation{
  $^1$Department of Physics, University of Tokyo, Tokyo, 113-0033, Japan \\
  $^2$JST, PRESTO, Tokyo, 113-0033, Japan \\
  $^3$Institute for Solid State Physics, University of Tokyo, Kashiwa, 277-8581, Japan
}

\maketitle
%!TEX encoding = UTF-8 Unicode
\section{BTRG for the honeycomb lattice}
\begin{figure}
  \begin{center}
    \includegraphics[clip, width=8cm]{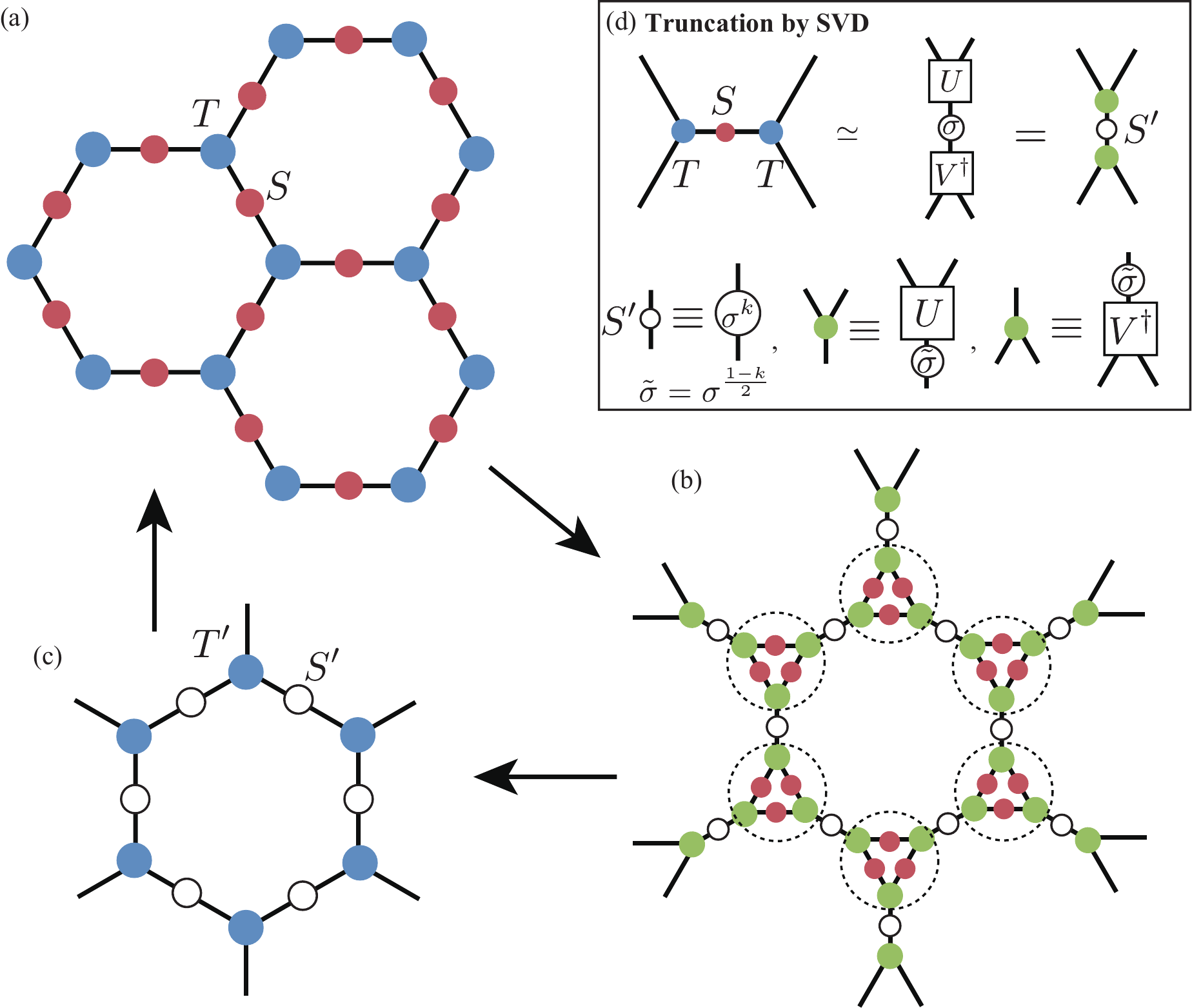}
  \end{center}
  \caption{  \label{BTRG_honeycomb}
    BTRG for the honeycomb-lattice Ising model. (a-c)~Renormalization steps for the honeycomb lattice. Similar to BTRG for the square lattice, we introduce bond tensors on the tensor network. (d)~Tensor decomposition and definition of new tensors. Unlike the conventional TRG, we consider SVD and low-rank approximation for the triplet of tensors, $TST$.
}
\end{figure}

In this section, we apply the BTRG idea to the honeycomb-lattice Ising model. As shown in Fig.~\ref{BTRG_honeycomb}, the renormalization step is similar to the original TRG~\cite{LevinN2007} except for the existence of the bond tensors. Unlike the conventional TRG, we consider SVD and the low-rank approximation for the triplet of tensors, $TST$. We introduce a hyperparameter $k$, by which we define the new bond tensor as $S' = \sigma^k$, where $\sigma$ is the singular value matrix obtained by SVD of the triplet.
\begin{figure}
  \begin{center}
    \includegraphics[clip, width=8cm]{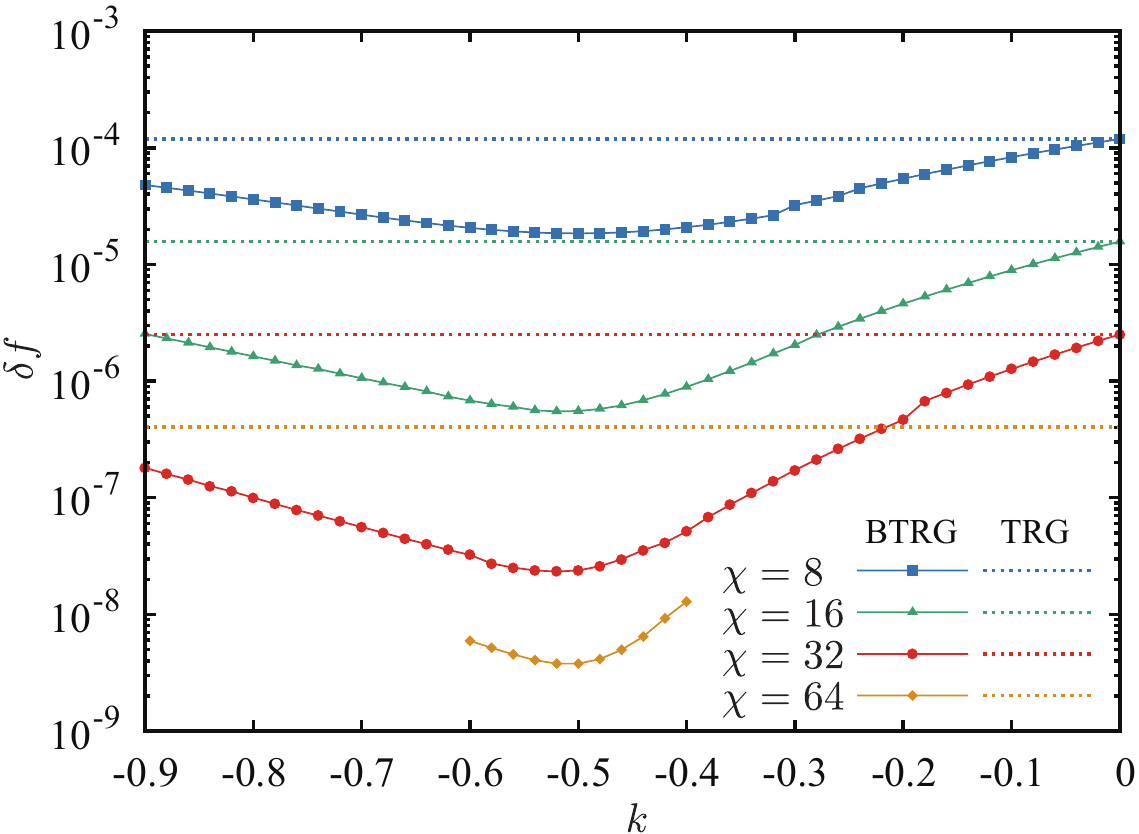}
  \end{center}
  \caption{  \label{BTRG_honeycomb_free_energy}
    $k$-dependence of the relative error of the free energy $\delta f = \lVert f_{\mathrm{calc}}-f_{\mathrm{exact}}\rVert / \lVert f_{\mathrm{exact}} \rVert$ obtained by BTRG at the critical point of the honeycomb-lattice Ising model for $\chi = 8, 16, 32$, and $64$. The horizontal dotted lines represent the results of TRG. $f_{\mathrm{calc}}$ is calculated for $N=3^{21}$ spins, and the $f_{\mathrm{exact}}$ is the free energy density in the thermodynamic limit.}
\end{figure}

In Fig.~\ref{BTRG_honeycomb_free_energy}, we show the $k$-dependence of the relative error of the free energy obtained by BTRG at the critical point $\beta_c=\frac{1}{2}\log(2+\sqrt{3})$ for  $\chi = 8, 16, 32$, and $64$. We can see that the relative error is largely reduced for negative $k$, and at $k \approx -0.5$ it shows the minimum, similar to the case of the square-lattice tensor network. Indeed, the stationary condition for the honeycomb-lattice tensor network is given by
\begin{equation}
\big[\sigma^{\frac{1-k}{2}}\big]^6 \big[\sigma^k\big]^7 = \sigma,
\end{equation}
where $\sigma^k$ corresponds to the bond-tensor $S$ and $\sigma^{\frac{1-k}{2}} = \tilde{\sigma}$ [see Fig.~\ref{BTRG_honeycomb}(d)]. From the stationary condition, we obtain $k=-\frac{1}{2}$ and it is consistent with our observation in Fig.~\ref{BTRG_honeycomb_free_energy}.

\section{Bond-weighted higher-order tensor renormalization group (BHOTRG)}

\begin{figure}
  \begin{center}
    \includegraphics[clip, width=8cm]{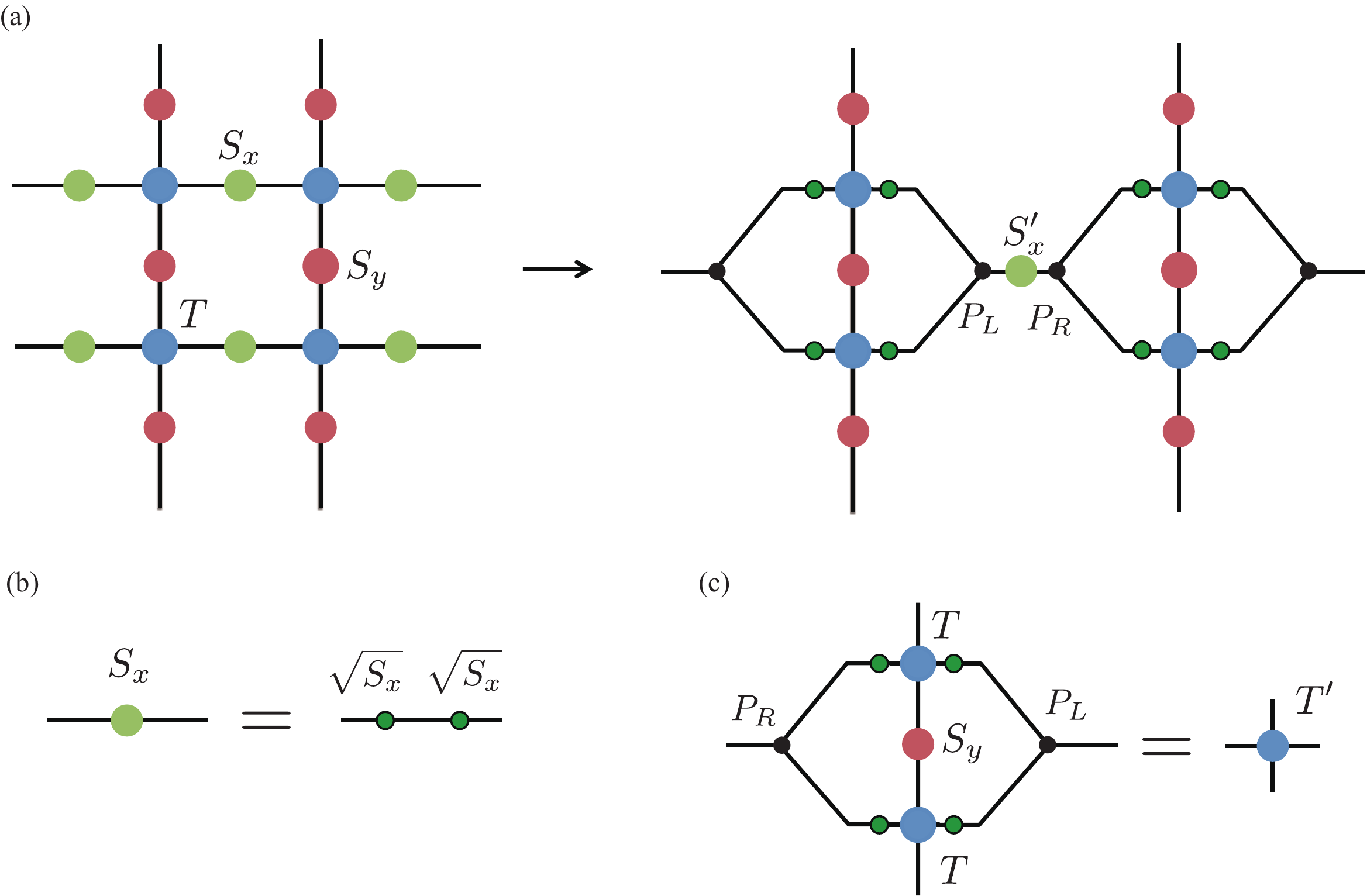}
  \end{center}
  \caption{  \label{BHOTRG}
    BHOTRG for the square-lattice Ising model. (a) Renormalization of two site tensors connected by a vertical bond. We insert a pair of squeezers $P_L$ and $P_R$ in order to truncate the bond degrees of freedom. (b) In BHOTRG, we split a bond tensor into two parts. (c) Definition of the new site tensor.
  }
\end{figure}

In this section, we briefly explain the application of our bond-weighted idea to HOTRG~\cite{XieCQZYX2012}. We call this approach as the bond-weighted HOTRG (BHOTRG). In BHOTRG, similar to BTRG, we consider tensor renormalization group for the square-lattice tensor network with site and bond tensors. 

\begin{figure}
  \begin{center}
    \includegraphics[clip, width=7cm]{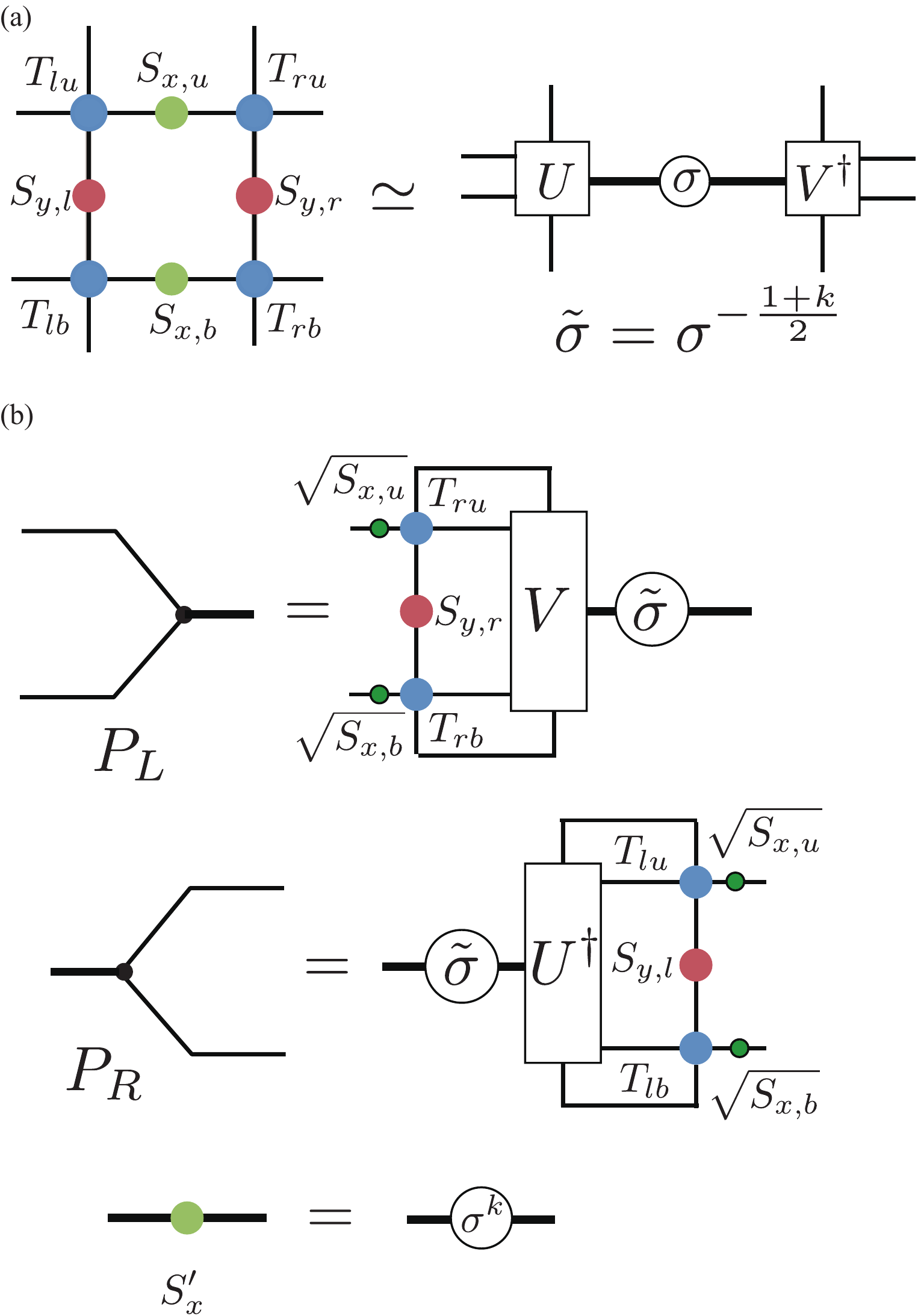}
  \end{center}
  \caption{  \label{BHOTRG_projector}
    Definition of the squeezers $P_L$ and $P_R$ in BHOTRG. (a) From the low-rank approximation by SVD, we obtain isometries $U$ and $V$ in addition to the singular value matrix $\sigma$. By introducing a hyperparameter $k$, $\tilde{\sigma}$ is defined as $\tilde{\sigma} = \sigma^{-(1+k)/2}$. (b) By using $U$, $V$, and $\tilde{\sigma}$, the squeezers, $P_L$ and $P_R$, and the renormalized bond weight $S_x'$ are defined.
}
\end{figure}

The outline of BHOTRG is shown in Fig.~\ref{BHOTRG}. As in the case of HOTRG, we perform renormalization along the horizontal and the vertical axes alternatively. The squeezers, $P_L$ and $P_R$, and the new bond weight $S_x'$ are defined as shown in Fig.~\ref{BHOTRG_projector}. As a natural generalization of HOTRG to inhomogeneous tensor networks, the squeezers are calculated from SVD of the diagram shown in Fig.~\ref{BHOTRG_projector}(a). Similar to BTRG, we introduce a hyperparameter $k$, which characterizes the difference from the conventional HOTRG. When we set the initial bond tensors as identity matrices, $S_x$ and $S_y$ remains as the identity matrices along the renormalization for $k=0$. In addition, when the initial site tensors have the ``mirror'' symmetry, such as the tensors in the partition function of the Ising model, the singular values in Fig.~\ref{BHOTRG} corresponds to those of HOSVD in the conventional HOTRG \cite{XieCQZYX2012}, and thus the BHOTRG with $k=0$ is reduced to HOTRG. 

\begin{figure}
  \begin{center}
    \includegraphics[clip, width=8cm]{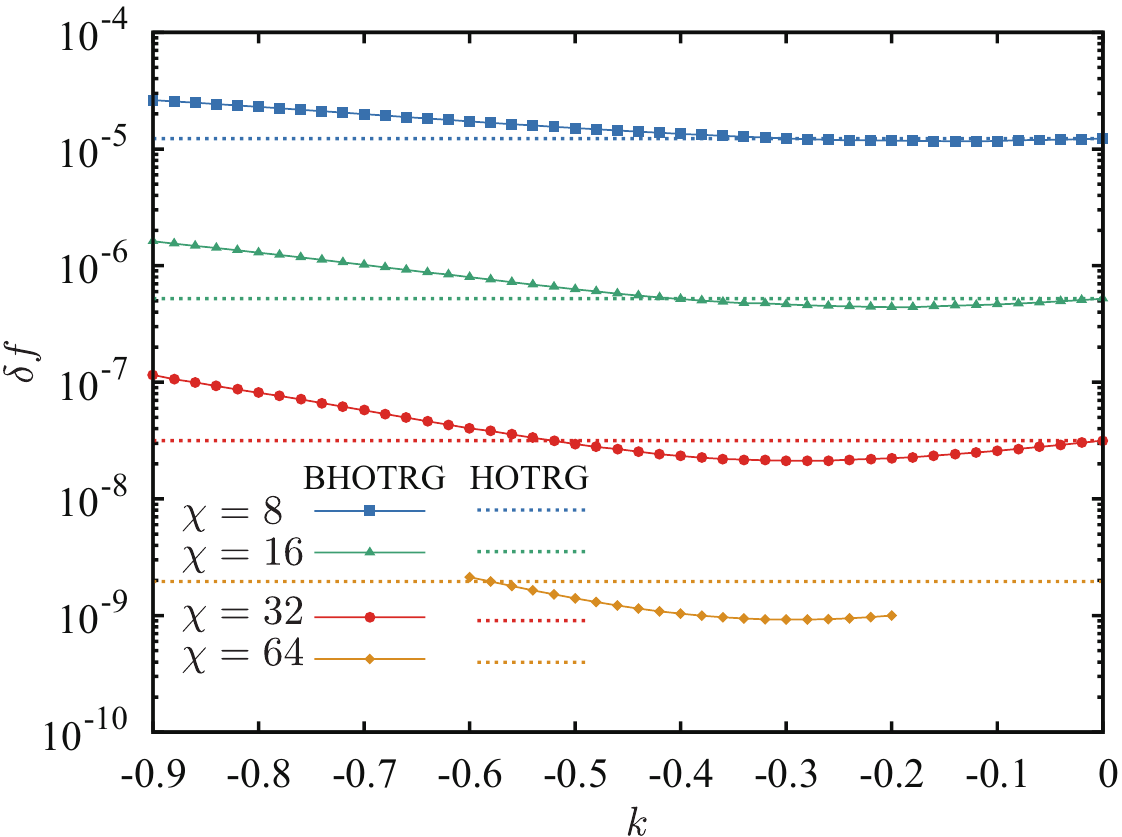}
  \end{center}
  \caption{  \label{BHOTRG_free_energy}
    $k$-dependence of the relative error of the free energy $\delta f = \lVert f_{\mathrm{calc}}-f_{\mathrm{exact}}\rVert / \lVert f_{\mathrm{exact}} \rVert$ obtained by BHOTRG at the critical point of the square-lattice Ising model for $\chi = 8, 16, 32$, and $64$. Horizontal dotted lines represent the results of HOTRG. $f_{\mathrm{calc}}$ is calculated for $N=2^{40}$ spins, and the $f_{\mathrm{exact}}$ is the free energy density in the thermodynamic limit.}
\end{figure}

In Fig.~\ref{BHOTRG_free_energy}, we show the $k$-dependence of the relative error of the free energy obtained from BHOTRG at the critical point of the square-lattice Ising model for $\chi=8$, 16, 32, and 64. Similar to the case of BTRG, we find better accuracy for nonzero $k$, though the accuracy is not better than BTRG (see also Fig.~2 in the main text%\ref{BTRG_result1}
). Also, the optimal value of $k$ shows a large dependence on $\chi$ compared with the case of BTRG. 

\begin{figure}
  \begin{center}
    \includegraphics[clip, width=8cm]{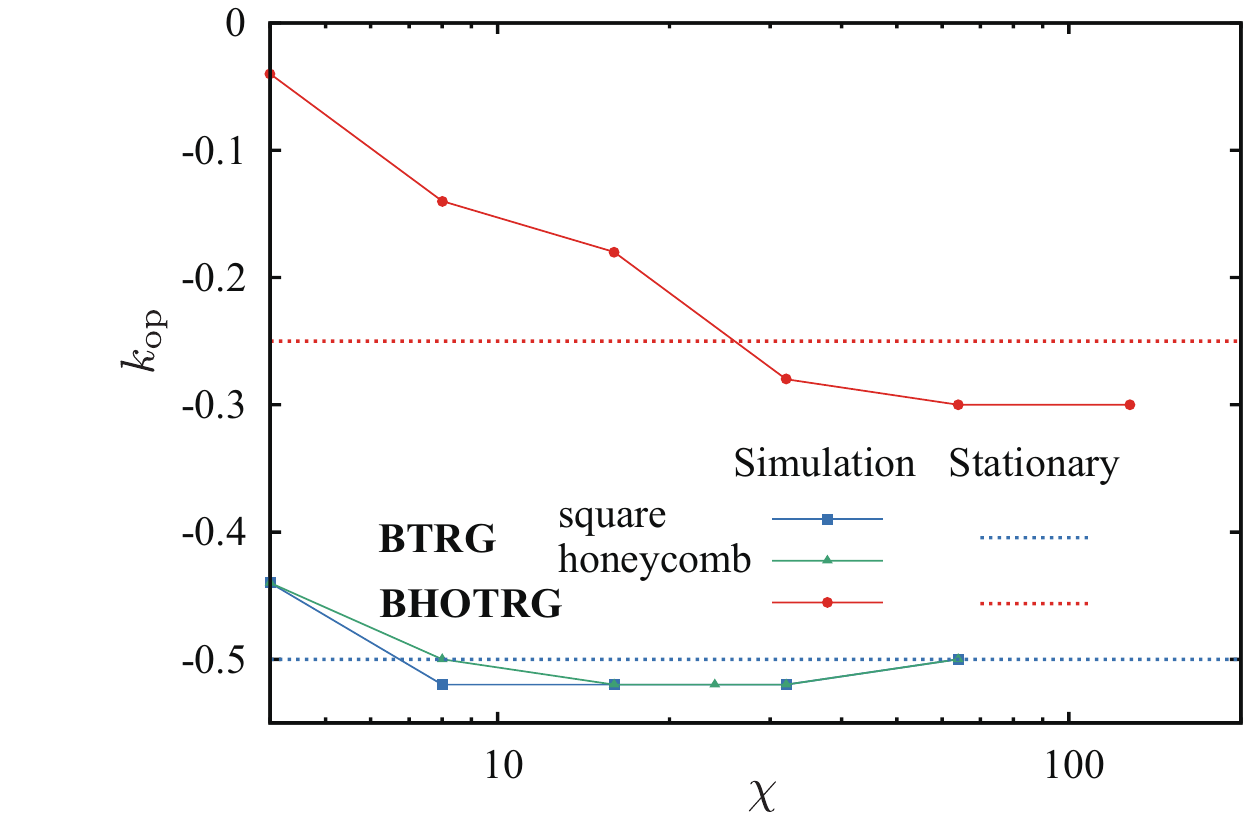}
  \end{center}
  \caption{  \label{Stationary_comp}
    $\chi$-dependence of the optimal value of $k$ obtained by BTRG and BHOTRG. Dotted lines represent the solutions of stationary conditions: $k = -\frac{1}{2}$ and $k=-\frac{1}{4}$ for BTRG and BHOTRG, respectively.}
\end{figure}

The stationary condition for BHOTRG can be obtained from Figs.~\ref{BHOTRG} and \ref{BHOTRG_projector} as 
\begin{align}
 T^4 \big[\sigma^{k}\big]^4 &= \sigma \notag \\
 T^6 \big[\sigma^{k}\big]^7 \big[\sigma^{-\frac{1+k}{2}}\big]^2 &= T,
\end{align}
where $T$ represents the weight of the site tensors, $\sigma^k$ corresponds to $S_x$ and $S_y$ and $\sigma^{-\frac{1+k}{2}} = \tilde{\sigma}$. From the above stationary condition, we obtain $k = -\frac{1}{4}$. In Fig.~\ref{Stationary_comp}, we show the optimal value of $k$ as a function of $\chi$ for BTRG and BHOTRG. We find that although the optimal $k$ does not converge to the solution of the stationary condition up to $\chi = 128$, it is not far from $k=-\frac{1}{4}$. One of the reason for such a large $\chi$-dependence of the optimal value might be the anisotropic renormalization in BHOTRG. In BHOTRG, the symmetry between the horizontal and the vertical directions are broken. Thus, the assumption $S_x = S_y$, which is used to derive the stationary condition, is not sufficiently fulfilled. This point also explain the small improvement of the relative errors in BHOTRG.

\bibliography{BTRG}